\documentclass[10pt]{bmc_article}

\usepackage{cite} % Make references as [1-4], not [1,2,3,4]
\usepackage{url}  % Formatting web addresses  
\usepackage{ifthen}  % Conditional 
\usepackage{multicol}   %Columns
\usepackage[utf8]{inputenc} %unicode support
\usepackage{amsmath}
\usepackage{amssymb}
\usepackage{amsbsy}

\usepackage{graphicx}

\newboolean{publ}

\newenvironment{bmcformat}{\begin{raggedright}\baselineskip20pt\sloppy\setboolean{publ}{false}}{\end{raggedright}\baselineskip20pt\sloppy}

\renewcommand{\matrix}[1]{\begin{pmatrix}#1\end{pmatrix}}

\newcommand{\T}{^\mathrm{T}}
\newcommand{\Real}{\mathbb{R}}
\newcommand{\R}{\mathbb{R}}
\newcommand{\N}{\mathbb{N}}

\newcommand{\setP}{\mathcal{P}}
\newcommand{\cP}{\mathcal{P}}

\newcommand{\param}{\theta}
\newcommand{\propensity}{\nu}

\newcommand{\stoich}{N}
\newcommand{\rsubstrate}{\sigma}
\newcommand{\rproduct}{\varphi}

\begin{document}
\begin{bmcformat}

\title{Efficient parametric analysis of the chemical master equation through model order reduction}
 
\author{Steffen Waldherr\correspondingauthor$^{1}$%
       \email{Steffen Waldherr\correspondingauthor - waldherr@ist.uni-stuttgart.de}%
      and
         Bernard Haasdonk$^2$%
         \email{Bernard Haasdonk - haasdonk@mathematik.uni-stuttgart.de}%
      }

\address{%
    \iid(1)Institute for Systems Theory and Automatic Control, University of Stuttgart, Pfaffenwaldring 9, %
         Stuttgart, Germany\\
    \iid(2)Institute for Applied Analysis and Numerical Simulation, University of Stuttgart, 
    Pfaffenwaldring 57, %
}%

\maketitle

\begin{abstract}
        % Do not use inserted blank lines (ie \\) until main body of text.
        \paragraph*{Background:} 
Stochastic biochemical reaction networks are commonly modelled by the chemical master
equation, and can be simulated as first order linear differential equations
through a finite state projection.
Due to the very high state space dimension of these equations, numerical simulations
are computationally expensive.
This is a particular problem for analysis tasks requiring repeated simulations for
different parameter values.
Such tasks are computationally expensive to the point of infeasibility with
the chemical master equation.
      
        \paragraph*{Results:} 
In this article, we apply parametric model order reduction techniques 
in order to construct accurate low-dimensional parametric models of the chemical master equation.
These surrogate models can be used in various parametric analysis task such as identifiability analysis, parameter estimation,
or sensitivity analysis.
As biological examples, we consider two models for gene regulation networks,
a bistable switch and a network displaying stochastic oscillations.

        \paragraph*{Conclusions:} 
The results show that the parametric model reduction yields efficient models of 
stochastic biochemical reaction networks, and that these models can be useful for
systems biology applications involving parametric analysis problems such 
as parameter exploration, optimization, estimation or sensitivity analysis.
\end{abstract}

\ifthenelse{\boolean{publ}}{\begin{multicols}{2}}{}

\section*{Background}

The chemical master equation (CME) is the most basic mathematical description of stochastic biomolecular reaction networks \cite{Gillespie1992,Kampen1981}.
The CME is a generally infinite-dimensional linear differential equation.
It characterizes the temporal development of the probabilities that the network is in any of its possible configurations, where the different configurations are characterized by the molecular copy numbers of the network's chemical species.

Due to its infinite dimension, the CME is usually not directly solvable, not even with numerical methods.
A recent breakthrough in the numerical treatment of the CME was the establishment of the finite state projection (FSP) method by Munsky and Khammash \cite{MunskyKha2006}.
They showed that it is possible to compute a good approximation to the real solution by projecting the CME to a suitable finite subdomain of the network's state space, and solving the resulting finite-dimensional linear differential equation on that domain.
Nevertheless, the FSP approach still yields very high-dimensional models which are computationally expensive to simulate, even for small biochemical networks.
The efficient simulation of the CME is an area of active research, and recently other simulation methods have been developed that can also be used for larger networks \cite{JahnkeHui2008a,HeglandHel2008}.

Despite this progress, the direct simulation of the CME remains a computational bottleneck for common model analysis tasks in systems biology.
It is especially problematic for tasks which require the repeated simulation of the model using different parameter values, for example identifiability analysis, parameter estimation, or model sensitivity analysis.
Thereby, while a single or a few evaluations of a CME model with the FSP or other approaches may still be computationally feasible, the necessity of many repeated simulations will quickly render higher-level analysis tasks infeasible.

Mathematical methods that approximate the behaviour of a high-dimensional original model through a low-dimensional reduced model are a common way to deal with complex models.
Especially for linear differential equations, model order reduction is a well established field and several methods to compute reduced order models are available \cite{Antoulas2005}.
Note that the step of generating a reduced model is usually computationally more expensive than a single or even a few simulations of the original high-dimensional model.
But the simulation of the resulting reduced models is frequently orders of magnitude faster than the solution of the original model. 
So, model reduction is worth the effort if many repeated 
simulations are to be expected.
Unfortunately, for analysis tasks which require the repeated model simulation with different parameters, classical model reduction methods are not helpful.
With these methods, the reduced model depends on specific parameter values in the original model, and the reduction needs to be redone for different parameter values.
Thus, for the mentioned analysis tasks, the model reduction process would have to be repeated for each new parameter value, and no gain in computational efficiency would typically be possible.
While classical model reduction techniques have been applied to the CME in the past \cite{MunskyKha2008a}, they are not so suitable for parametric analysis tasks.

Fortunately, model reduction methods where parameters from the original model are retained as adjustable parameters also in the reduced model are now being developed.
These methods allow to compute a reduced model which uses the same parameters as the original model, and where the reduced model can directly be simulated with any choice of parameter values 
\cite{BaurBenner2008,HaasdonkOhlberger2010,DSCLW04,MRGKH05}.

The purpose of this paper is to introduce the application of these parametric model reduction methods to finite-state approximations of the chemical master equation, and to show possible usage scenarios of such an approach.
The structure is as follows.
In the following section, we introduce some background and notation concerning the modelling of 
chemical reaction networks and parametric model order reduction. 
We also show how the parametric model order reduction methods can in fact be applied to the CME.
Afterwards, we apply the reduction technique on
two reaction network models and corresponding parametric analysis tasks.

\section*{Theoretical basics and methods}
\label{sec:preliminaries}

We start with some preparatory background on the chemical master equation (CME) and
parametric model order reduction. This serves in particular to fix the 
notation used throughout the remainder of the article.
Then the application of parametric model order reduction to the CME is introduced.

\subsection*{The chemical master equation}

The structure of a biochemical reaction network is characterized completely by the list of involved species, denoted as $X_1, X_2 \dotsc, X_n$, and the list of reactions, denoted as
\begin{equation}
\label{eq:brn-reactions}
\begin{aligned}
\sum_{i=1}^n \rsubstrate_{ij} X_i \rightarrow \sum_{i=1}^n \rproduct_{ij} X_i,\quad j=1,\ldots,m,
\end{aligned}
\end{equation}
where $m$ is the number of reactions in the network, and the factors $\rsubstrate_{ij} \in \mathbb{N}_0$ and $\rproduct_{ij} \in \mathbb{N}_0$  are the stoichiometric coefficients of the reactant and product species, respectively \cite{Higham2008}.
The net change in the amount of species $i$ occuring through reaction $j$ is given by
\begin{equation}
\begin{aligned}
\stoich_{ij} &= \rproduct_{ij} - \rsubstrate_{ij}.
\end{aligned}
\end{equation}
Reversible reactions can always be written in the form \eqref{eq:brn-reactions} by splitting the forward and reverse path into two separate irreversible reactions.

For a stochastic network model, the variables of interest are the probabilities that the network is in any of the possible states which are characterized by the molecular copy numbers of the individual species $X_1, X_2 \dotsc, X_n$.
We denote the molecular copy number of $X_i$ by $[X_i] \in \mathbb{N}_0$.
Then, the state variables of the stochastic model are given by the real numbers
\begin{equation}
\label{eq:state-variable-p}
\begin{aligned}
p(t,x_1,x_2,\dotsc,x_n) = \mathrm{Prob}([X_1] = x_1, [X_2] = x_2, \dotsc, [X_n] = x_n \textnormal{ at time } t),
\end{aligned}
\end{equation}
for $x_i \in \mathbb{N}_0$, $i=1,\dotsc,n$.
As a short-hand notation for \eqref{eq:state-variable-p}, we write $p(t,x)$, with $x \in \mathbb{N}_0^n$.

The transitions from one state to another are determined by chemical reactions according to \eqref{eq:brn-reactions}.
The changes in the molecule numbers are described by the stoichiometric reaction vectors
\begin{equation}
\begin{aligned}
v_j = \matrix{\stoich_{1j} & \stoich_{2j} & \dotsb & \stoich_{nj}}\T \in \mathbb{Z}^n.
\end{aligned}
\end{equation}
% where each element $\stoich_{ij}$ corresponds to the change in the molecular number $x_i$ occuring under reaction $j$.
To avoid needlessly complicated cases, we assume $v_j \neq v_k$ for $j \neq k$.

The probabilities of the network being in any of the possible states $x$ evolve over time, and their evolution is governed by the chemical master equation (CME) as derived by \cite{Gillespie1992}.
From a given molecular state $x$, one can compute the propensity $\propensity_j$ that reaction $j$ takes place according to the law of mass action as
\begin{equation}
\label{eq:reaction-propensity}
\begin{aligned}
\propensity_j(x,\param) = \param_j \prod_{i=1}^n {x_{i} \choose \rsubstrate_{ij}},
\end{aligned}
\end{equation}
where $\param=(\param_j)_{j=1}^m$ is the vector of reaction rate 
constants, which are model parameters depending on the physical properties of 
the molecules involved in the reactions.
The propensities are related to the probability that reaction $j$ will occur in a short time interval of length $dt$ when the system is in state $x$:
\begin{equation}
\label{eq:reaction-probability}
\begin{aligned}
\mathrm{Prob}(\textnormal{reaction $j$ occurs in $[t,t+dt]$} \mid [X] = x) = \propensity_j(x,\param) dt + {\scriptstyle\mathcal{O}}(dt).
\end{aligned}
\end{equation}

Taking the possible transitions and the corresponding reaction propensities together yields the chemical master equation (CME), a linear differential equation where the variables are the probabilities that the system is in each of the possible molecular states $x$:
\begin{equation}
\label{eq:cme}
\begin{aligned}
\frac{d}{dt} p(t,x) = \sum_{j=1}^m (\propensity_j(x-v_j, \param) p(t,x-v_j) - \propensity_j(x,\param) p(t,x)),
\end{aligned}
\end{equation}
for $x \in \mathbb{N}_0^n$.
The CME \eqref{eq:cme} is subject to an initial condition 
$ p(t_0,x) =  p_0(x)$ 
for $x\in\mathbb{N}_0^n$.

Despite being linear, the CME is hard to solve numerically.
This is due to the problem that the state space is for most systems infinite-dimensional, since all possible states $x \in \mathbb{N}_0^n$ of the reaction network \eqref{eq:brn-reactions} must in general be considered.
Instead of directly solving the CME \eqref{eq:cme}, a number of alternative approaches to study the stochastic dynamics of biochemical reaction networks have been suggested.
The most common approach is to generate a simulated realization of the stochastic process described by the reaction network \eqref{eq:brn-reactions}, using for example the Gillespie algorithm \cite{Gillespie1977}.
In this approach, the probabilities $p(t,x)$ for the possible system states are obtained from many simulated realizations.
However, since this requires a large number of realizations, it is computationally expensive.

As a more direct approach, Munsky and Khammash \cite{MunskyKha2006} have proposed the finite state projection (FSP), where the CME is solved on a finite subset of the state space.
Here, this subset is denoted by $\Omega$, and is defined as
\begin{equation}
\label{eq:omega}
\Omega = \lbrace x^{(i)} \mid i = 1,\dotsc,d \rbrace \subset \mathbb{N}_0^n,
\end{equation}
where the $x^{(i)}$ are the system states for which the probabilities are computed in the projected model.
The underlying assumption is that the probabilities for other states will be very low on the time scale of interest---otherwise the FSP may not yield good approximations to the solution of the CME. 
In particular we assume the time interval of interest to be given 
by $[0,T]$ for final time $T>0$.
The probabilities for the states $x^{(i)}$ in $\Omega$ are written 
in the vector $P(t)$ approximating $p(x,t)$ at the finite number of 
states $\Omega$:
\begin{equation}
  \label{eq:vector-P}
  P(t) = \bigl( P_i(t)\bigr)_{i=1,\dotsc,d} \approx \bigl( p(t,x^{(i)}) \bigr)_{i=1,\dotsc,d} \in [0,1]^d.
\end{equation}
The equation to be solved with the FSP approximation is
\begin{equation}
  \label{eq:cme-finite}
  \begin{aligned}
    \frac{d}{dt} P(t) &= A(\param) P(t) \quad \mbox{ for } t\in (0,T)\\
    P(0) &= P_0,
  \end{aligned}
\end{equation}
where $A(\param) \in \Real^{d\times d}$ is the matrix of state transition propensities, and $P_0 = \bigl( p_0(x^{(i)}) \bigr)_{i=1,\dotsc,d}$ is a vector of initial probabilities for the states in $\Omega$.
The elements of the matrix $A(\param)$ are computed as
\begin{equation}
  \label{eq:elements-A}
  \begin{aligned}
    A_{ii}(\param) &= - \sum_{r=1}^m \propensity_r(x^{(i)},\param) \\
    A_{ij}(\param) &= \left\lbrace
    \begin{aligned}
      \propensity_r(x^{(j)},\param) & \textnormal{ if } x^{(j)} = x^{(i)} + v_r,\ r = 1,\dotsc,m \\
      0 & \textnormal{ otherwise}.
    \end{aligned}
    \right.
  \end{aligned}
\end{equation}
We will frequently omit the parameter dependence of the solution 
(and other parametric quantities). Hence the
solution $P(t)$, as abbreviation of $P(t,\param)$,
of \eqref{eq:cme-finite} is an 
approximation to the solution $p(t,x)$ of the orginal CME on the 
domain $\Omega$.
Munsky and Khammash \cite{MunskyKha2006} have also derived an upper bound on the error between the solution $P(t)$ computed via the FSP, and the solution of the original CME $p(t,x)$ on $\Omega$.

Here, we consider in addition an output vector $y\in\Real^p$ defined by
\begin{equation}
\label{eq:output-equation}
y(t) = C P(t),
\end{equation}
with $C \in \Real^{p\times d}$.
Examples for relevant outputs are the probability that the molecular copy numbers are in a certain domain $\bar\Omega \subset \Omega$, which is achieved by the row vector output matrix $C$ defined by $C_i = 1$ if $x^{(i)} \in \bar\Omega$, otherwise $C_i = 0$, with $p=1$, or the expected molecular copy numbers, given by
\begin{equation}
  \label{eq:output-average}
  y_{e}(t) = \sum_{i=1}^d x^{(i)} P_i(t),
\end{equation}
i.e.\ $C = (x^{(1)}, \dotsc, x^{(d)})$ with $p=n$.

The basic motivation for the model reduction presented here is that we are interested in parametric analysis of the model, where the model \eqref{eq:cme-finite} has to be solved many times with different values for the parameters $\param$.
Due to the typical high dimensions of the matrix $A(\param)$, already a single simulation is computationally expensive, and analysis tasks requiring many repeated simulations are often computationally infeasible.
%not so much interested in the actual state vector $P$, but rather in the output vector $y$.
Thus, the primary goal is to derive a reduced model which is rapidly 
solvable and provides an approximation $\hat y(t)$ to the output $y(t)$, potentially 
without any consideration of the original state vector $P(t)$.

\subsection*{Order reduction of parametric models}
\label{sec:order-reduction-background}

Model order reduction of parametric problems is a very active research field in systems theory, engineering and applied mathematics. 
We refer to \cite{BaurBenner2008,DSCLW04,MRGKH05} and references 
therein for more information on the topic. 

Here, we apply the reduction technique for parametric problems presented in
\cite{HaasdonkOhlberger2010} adopted to our notation. 
It is based on two biorthogonal global projection 
matrices $V,W\in \R^{d \times r}$ with $r\ll d$ and $W^T V = Id$, where $r$ is the dimension of the reduced model.
The matrix $V$ is assumed to span a space that approximates the system state 
variation for all parameters and times. 
The construction of such matrices will be detailed in 
the next subsection. 

The gain of computational efficiency in repeated simulations comes from a separation of the simulation task into a computationally expensive ``offline'' phase and a computationally cheap ``online'' phase.
In the offline phase, suitable projection matrices $V$ and $W$ are computed without fixing specific parameter values.
With the projection matrices, a reduced model with the same free parameters as the original model is computed.
In the online phase, the reduced model is simulated with the actually chosen parameter values, which is typically several orders of magnitude faster than the simulation of the original model.
For analysis tasks with repeated simulations, only the online phase has to be repeated for different choices of the parameter values, yielding an overall gain in computational efficiency.

\subsubsection*{Decomposition in parametric and non-parametric part}
\label{sec:decomposition}

The reduction technique assumes a separable parameter dependence of the 
full system matrices and the initial condition. This means, we assume that 
there exist a suitable small constant $Q_A\in \N$, parameter independent 
components $A^{[q]} \in \R^{d \times d}$ and parameter dependent scalar 
coefficient functions $\vartheta^{[q]}_A(\param)$ for $q=1,\ldots,Q_A$ 
such that
\begin{equation}
  \label{eq:decomposition}
  A(\param) = \sum_{q=1}^{Q_A} \vartheta^{[q]}_A(\param) A^{[q]}
\end{equation}
and similarly for the system matrix $C$ and initial condition $P_0$.
We assume that $\param\subset \cP$ stems 
from some domain $\cP \subset \R^{m}$ of admissible parameters.
In the next step, the reduced component 
matrices and initial conditions are determined by
\begin{equation}
  A_r^{[q]} := W^T A^{[q]} V, \quad  C_r^{[q]} := C^{[q]} V, \quad 
 P_{0r}^{[q]}:= W^T P_0^{[q]}. 
\end{equation}
for $q = 1, \dotsc, Q_A$.
The resulting quantities $A_r^{[q]}$, $C_r^{[q]}$, and $P_{0r}^{[q]}$ are $r$-dimensional vectors or matrices 
and independent of the high dimension $d$.
The basis computation and the computation of these reduced system components is performed once and parameter-independently in the offline-phase.
Then, in the online-phase, for any new parameter $\param$ the reduced system 
matrices and the initial condition are assembled by 
\begin{equation}
A_r(\param) = \sum_{q=1}^{Q_A}\vartheta^{[q]}(\param)A_r^{[q]}
\end{equation}
and similarly
for $P_{r0}(\param)$ and $C_r(\param)$. 
The low dimensional reduced system that remains to be solved is 
\begin{equation}
  \label{eq:cme-finite-reduced}
  \begin{aligned}
    \frac{d}{dt} P_r(t) &= A_r(\param) P_r(t) \quad \mbox{ for } t \in (0,T)\\
    P_r(0) &= P_{r0}(\param)\\
    \hat y(t) &= C_r(\param) P_r(t).
  \end{aligned}
\end{equation}
From the reduced state $P_r(t)$, an approximate state for the full system can be reconstructed at any desired time 
by $\hat P(t)= V P_r(t)$.
Also the difference between the approximated output $\hat y(t)$ and the output $y(t)$ of the original model can be bounded by so called error estimators.
A-posteriori error bounds for the reduced systems as considered here are given in
\cite{HaasdonkOhlberger2010}.

\subsubsection*{Basis generation}
\label{sec:basis-generation-background}

Different methods for the computation of the projection bases $V$ and $W$ exist.
In systems theory, methods like 
balanced truncation, Hankel-norm approximation or moment matching 
are applied, that approximate the input-output behaviour of a linear 
time-invariant system \cite{Antoulas2005}. 
The resulting reduced models can be applied for 
varying input signals. 
Extensions to parametric problems exist, e.g. 
\cite{BaurBenner2008,MRGKH05}.
As we do not have varying inputs in the problem studied here, we consider snapshot-based approaches to be
more suitable. This means, the projection bases are constructed by 
solution snapshots, i.e. special solutions computed for 
selected parameter values.

%\subsubsection*{Balanced Truncation}
%balanced truncation
%
%conditions for balanced truncation \cite{Antoulas2005}:
%\begin{itemize}
%\item $A$ stable
%\item system reachable
%\item system observable
%\end{itemize}

%\subsubsection*{POD-Greedy}

The generation of projection matrices $V$ and $W$ must be done in such 
a way, that they are globally well approximating the system states over the 
parameter and time domain. 
A possible way to achieve this is the POD-Greedy algorithm, 
which has been introduced in 
\cite{HO08a} and
is meanwhile a standard procedure in reduced basis methods 
\cite{EKP11,KP10}. 
The algorithm makes use of a repeated proper orthogonal decomposition (POD)
of trajectories $P:[0,T]\rightarrow \R^d$, which for our purposes can be defined as 
\begin{equation}
  \label{eq:pod}
POD(P) := \arg \min_{v \in \R^d,||v||=1} \int_{0}^T  
  || P(t)- (v^T P(t)) v ||^2 dt.
\end{equation}
Intuitively, $POD(P) \in \R^d$ is a state space vector representing the
single dominant mode that minimizes the squared mean projection error.
Computationally, this minimization task is solved by a reformulation as a 
suitable eigenvalue problem. 
Consider the correlation matrix $C = \int_{0}^T  P(t) P(t)^T dt$.
Then, $v^\ast = POD(P) \in \R^d$ is an eigenvector corresponding to the largest eigenvalue $\lambda_{max}$ of $C$, i.e., $C v^\ast = \lambda_{max} v^\ast$.
For additional theoretical and computational details on POD we refer to \cite{Vo11,Jo:02}.
We further require a finite subset of parameters $\setP_{train}\subset \setP$, 
that are used in the basis generation process.
As error indicator $\Delta(\param,V)$ we use the projection 
error of the full system trajectory on the reduced space spanned by the 
orthonormal columns of $V$, i.e. 
\begin{equation}
  \label{eq:pod-error}
  \Delta(\param,V) := \int_{0}^T ||P(t,\param)- V V^T P(t,\param)||^2 dt.
\end{equation}
The POD-Greedy procedure which is given in the pseudo-code below, starts 
with an arbitrary orthonormal initial 
basis $V_{N_0}\in \R^{d \times N_0}$ 
and performs an incremental basis extension.
The algorithm repeatedly identifies the currently worst resolved parameter (a),
orthogonalizes the corresponding full trajectory with the current reduced 
space (b), computes a POD of the error trajectory (c), and inserts the dominant mode 
into the basis (d).

{
\vspace{0.2cm}
\noindent
{function} $V = $ POD-Greedy$(\setP_{train},V_{N_0},\varepsilon_{tol})$
\begin{enumerate}
\item $N:= N_0$
\item while 
$\varepsilon_N:=\max_{\param \in \setP_{train}} \Delta(\param,V_N)> \varepsilon_{tol}$
\begin{enumerate}
\item $\param^\ast := \arg \max_{\param \in \setP_{train}} \Delta(\param,V_N)$ 
\item $E(t) := P(t,\param) - V_N (V_N^T P(t,\param^\ast))$
\item $v_{N+1}:= POD(E)$
\item $V_{N+1} := [V_N, v_{N+1}]$
\item $N := N+1$
\end{enumerate}
\item end while
\end{enumerate}
}

Note that the algorithm is implemented such that the simulation of the full model, yielding 
$P(t,\param)$ in \eqref{eq:pod-error}, is only performed once for each $\param$ 
in the training set $\setP_{train}$.

For concluding the basis generation, we set $W:=V$. This satisfies the 
biorthogonality condition $W^T V=Id$,
as $V$ has orthonormal columns by construction. 
In practice the time-integrals in \eqref{eq:pod} are realized by a finite sampling of 
the time interval.

A theoretical underpinning for the POD-Greedy algorithm has recently been 
provided by the analysis of convergence rates \cite{Haasdonk2011}.
This is based on the approximation-theoretical 
notion of the {\em Kolmogorov $n$-width $d_N({\cal F})$} of a given set ${\cal F} \subset \Real^d$, which 
quantifies how well the set can be approximated by arbitrary $N$-dimensional linear subspaces of $\Real^d$.
The convergence statement for the case of exponential convergence then can be summarized as follows:
If the set of 
solutions ${\cal F}:=\{P(t,\param)| t\in [0,T], \param \in \cP\} \subset \R^d$ 
is compact and has an exponentially decaying Kolmogorov $n$-width
$d_N({\cal F}) \leq M e^{-aN^\alpha}$ for some $ M,a,\alpha>0$ and all $ N\in \N$, then 
the error sequence $(\varepsilon_N)_{N\in \N}$ 
generated by the POD-Greedy procedure (cf. the definition in Step 2.\ in the pseudo code) also 
decays with an exponential rate, 
$\varepsilon_N \leq CMe^{-cN^\beta}$ 
with suitable constants $\beta, c, C> 0 $ depending on $M,a,\alpha$. 
Thus, if the set of solutions can be approximated by linear subspaces with an exponentially decaying error term, then the POD-Greedy algorithm will in fact find an approximation with an exponentially decaying error term, though possibly with suboptimal parameters in the error bound.
% This error convergence behaviour is observed in our experiments.

Extensions of the POD-Greedy algorithm exist, e.g. allowing more than one 
mode per extension step, performing adaptive parameter and time-interval partitioning, or
enabling training-set adaptation \cite{EKP11,KP10,HDO10}.
%(potentially different bases in different regions of the parameter space)

\subsection*{Reduced models of the parametrized chemical master equation}
\label{sec:cme-model-reduction}

In this section, we describe how to apply the reduction method for parametrized models presented in the previous section to FSP models for the chemical master equation.

% \subsection*{Derivation of the reduced order model}
% \label{sec:decomposition}

As discussed in the previous section, the first step in the proposed reduction method is a decomposition of the $d$-dimensional system matrix $A(\param)$ as in \eqref{eq:decomposition}.
Such a decomposition is possible for the case of mass action reaction propensities, as defined in \eqref{eq:reaction-propensity}, or generalized mass action, as recently suggested for the chemical master equation \cite{WuVid2011}.
In this case, the length of the parameter vector $\param$ is equal to the number of reactions $m$, and we decompose $A(\param)$ into $m$ terms as
\begin{equation}
  \label{eq:decomposition-cme}
  A(\param) = \param_1 A^{[1]} + \dotsm + \param_m A^{[m]}.
\end{equation}
Hence, concerning the notation given before, we have $Q_A = m$ components 
$A^{[q]}$ and coefficient functions $\vartheta_A^{[q]}(\param)=\param_q$.
Each matrix $A^{[q]}$ in this decomposition comes from just the transition propensities corresponding to reaction $q$, and is defined by
\begin{equation}
  \label{eq:elements-decomposed-matrix}
  \begin{aligned}
    A^{[q]}_{ii} &= - \prod_{k=1}^n (x^{(i)}_{k})^{\rsubstrate_{kq}} \\
    A^{[q]}_{ij} &= \left\lbrace
    \begin{aligned}
      \prod_{k=1}^n (x^{(j)}_{k})^{\rsubstrate_{kq}} & \textnormal{ if } x^{(j)} = x^{(i)} + v_q \\
      0 & \textnormal{ otherwise}.
    \end{aligned}
    \right.
  \end{aligned}
\end{equation}
More generally, such a decomposition is also possible if reaction rate propensities can be decomposed into the product of two terms, with the first term depending on parameters only, and the second term on molecule numbers only.
This case is for example encountered when the temperature-dependance of the reaction rate constant is relevant, and the temperature $T$ is a variable parameter in the Arrhenius equation $\param = A e^{\frac{-E_A}{R T}}$. 
Since the output matrix $C$ and the initial condition $P_0$ are usually not depending on parameters in this framework, a decomposition of $C$ and $P_0$ is not considered.

The situation is more difficult for reaction propensities involving for example rational terms with parameters in the denominator.
The denominator parameters can not be included in the reduced order model by the decomposition outlined in \eqref{eq:decomposition-cme} and \eqref{eq:elements-decomposed-matrix}.
If variations in these parameters are however not relevant to the planned analysis, then they can be set to their nominal value, and the decomposition can directly be done as described above.
Alternatively, approximation steps can be performed, such as
Taylor series expansion or empirical interpolation \cite{BMNP04}, that 
generate an approximating parameter-separable expansion.

\section*{Results for exemplary applications in genetic switching and oscillations}
\label{sec:example-models}

In this section, we present the study of two example networks with the proposed model reduction method.
With these examples, the applicability of the reduced modeling approach especially for analysis tasks requiring repeated simulations with different parameter values is illustrated.
The first network is a bistable genetic toggle switch, where cells may switch randomly between two states, based on the model in \cite{WaldherrWu2010}.
For this network, the problem of parameter estimation with a reduced model is studied.
The second network is a second-order genetic oscillator, based on \cite{El-SamadKha2006a}, where we perform a sensitivity analysis over a wide parameter range.

\subsection*{Parameter estimation in a genetic toggle switch model}
\label{sec:genetic-switch-model}

\subsubsection*{Network description}
\label{sec:switch-network-description}

The genetic toggle switch considered here is an ovarian follicle switch model from \cite{WaldherrWu2010}.
It is a system of two genes which activate each other.
The switch is modelled as a reaction network with two species $X_1$, $X_2$, representing the gene products.
The network reactions are specified in Table~\ref{tab:switch-model}.

\begin{table}
  \centering
  \caption{List of reactions and reaction propensity functions for the follicle switch model \cite{WaldherrWu2010}.
    Nominal parameter values are
    $k_1 = 4$, 
    $V_1 = 75$,
    $M_1 = 25$,
    $u_1 = 0.01 \frac{1}{\mathrm{min}}$, 
    $V_2 = 75$,
    $M_2 = 25$,
    $u_2 = 0.01 \frac{1}{\mathrm{min}}$.}
  \label{tab:switch-model}
  \begin{tabular}{ccc}
    Reaction & Stoichiometry $v_j$ & Propensity $\propensity_j$ \\
    Production of $X_1$ & $(1, 0)\T$ & $u_1 (k_1 + \frac{V_1 x_2^3}{M_1^3 + x_2^3})$ \\
    Degradation of $X_1$ & $(-1, 0)\T$ & $u_1 x_1$ \\
    Production of $X_2$ & $(0, 1)\T$ & $u_2 (\frac{V_2 x_2^3}{M_2^3 + x_2^3})$ \\
    Degradation of $X_2$ & $(0, -1)\T$ & $u_2 x_2$ \\
  \end{tabular}
\end{table}

In \cite{WaldherrWu2010}, this network was shown to describe a bistable switch with two probability peaks, one close to $x^{(off)} = (0,0)\T$ and the other close to $x^{(on)} = (V_1, V_2)\T$.

In the study \cite{WaldherrWu2010}, only the lower probability peak was of interest.
Here, we are interested in
%, since the particular focus of this study was 
the transition of the system from $x^{(off)}$ to $x^{(on)}$.
Therefore, the system is truncated to a rectangle 
$\bar \Omega:= \{0,\ldots,150\}\times\{0,\ldots,150\}$ such that 
$x^{(on)},x^{(off)}\in \bar \Omega$, yielding a good approximation in the finite state projection to the infinite-dimensional chemical master equation.

%triangle with corner points $(0,0)$, $(\frac{V_1}{u_1}, 0)\T$, and $(0, \frac{V_2}{u_2})\T$, plus one single state representing the truncated part.

The next step is to apply the decomposition of the matrix $A(\param)$ as described in the methods section.
Note that $A(\param)$ for the switch network contains rational terms with the parameters $M_1$ and $M_2$.
Considering these two parameters as fixed quantities, the truncated CME for the follicle switch can be written as
\begin{equation}
  \label{eq:switch-detailed-ode}
  \dot P(t) = (k_1 A^{[1]} + V_1 A^{[2]} + u_1 A^{[3]} + V_2 A^{[4]} + u_2 A^{[5]}) P(t),
\end{equation}
where $A^{[i]}$, $i=1,\dotsc,5$ are of 
dimension 
$151^2 \times 151^2 = 22801 \times 22801$.
%$(\frac{V_1}{u_1}+1) (\frac{V_2}{u_2}+1) = 151^2 = 22801$ 
%$\frac{1}{2} (\frac{V_1}{u_1}+1) (\frac{V_2}{u_2}+1) = 1300$ 

%The initial condition suggested in \cite{WaldherrWu2010} is a probability of 1 for $x^{(off)}$.
%However, this initial condition turned out to be numerically ill-conditioned for the model reduction, so 
As initial condition we choose a probability distributed over some lower states
\begin{equation}
  \label{eq:switch-ic}
  p(0,x) = \left\lbrace
    \begin{aligned}
      \frac{1}{210} 
    &\quad \textnormal{for } x_1+x_2 \leq 20 \\
      0 &\quad \textnormal{otherwise.}
    \end{aligned}\right.
  \end{equation}
For the parametric model reduction, we consider only variations in the 
parameters $u_1$ and $u_2$.
These influence both the steady state level of gene activity in the on-state 
as well as the switching kinetics and are thus of high biological 
significance in the model. 
Hence we set $\param:=(u_1,\ u_2)^T \in [0.005,\ 0.02]^2$ as 
the parametric domain $\setP$.
As final time we choose $T=10^7$ which corresponds to a 
time range of approximately 19 years, i.e.\ about three times the half-life time of the off-state estimated in \cite{WaldherrWu2010}.

Some state plots from the simulation of the full model are shown in Figure \ref{fig:state-plots}. 
These snapshots clearly show the transition of the switch from the off-state with low values for $x_1$ and $x_2$ to the on state with high values.
The parameter influence is mainly reflected in the speed of the transition: for the parameter vector $(u_1,\ u_2) = (0.005,\ 0.02)$ in the lower row, most of the probability is already arranged around the on-state at the end of the simulation time.
In contrast, for the parameter vector $(u_1,\ u_2) = (0.05,\ 0.005)$ in the upper row, a significant portion of the probability is still located around the off-state at this time point.
Also, the transition paths are different: in the first case, the values for $x_2$ are lower than the values for $x_1$ during the transition, while in the second case, this relation is reversed.

As typical simulation time for a single trajectory of the full system, 
we obtain 98.2 seconds on a IBM Lenovo 2.53 GHz Dual Core Laptop.

\begin{figure}
  \centering
  \includegraphics{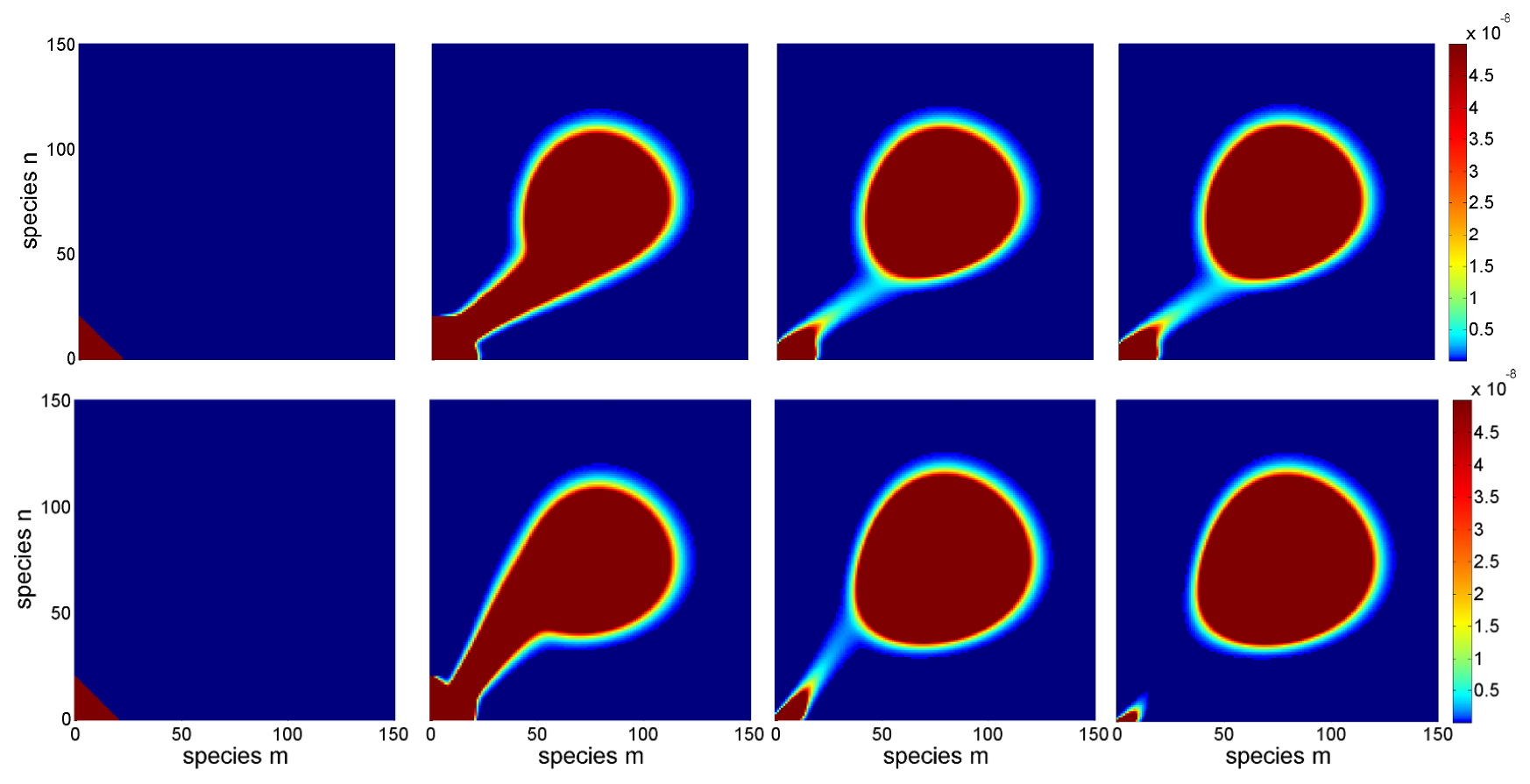}  
\caption{Illustration of some solution snapshots $P(t)$ of the switch 
model \eqref{eq:switch-detailed-ode} for parameter vector 
$(u_1,u_2) = (0.05, 0.005)$ (upper row) and
$(u_1,u_2) = (0.005, 0.02)$ (lower row) 
at times $t=0$, $2\cdot 10^5$, $5 \cdot 10^6$, and $1 \cdot 10^7$ from left to right.}
\label{fig:state-plots}
\end{figure}

\subsubsection*{Basis generation}
\label{sec:switch-basis-cme}

We generated a reduced basis with the POD-Greedy algorithm, where the training set was chosen as the vertices of a mesh with $9^2$ logarithmically equidistant parameter values over the parameter domain $\setP$.
We set $\varepsilon_{tol}=10^{-12}$ as target accuracy.
We use the projection error as error measure, hence precompute 
the 81 trajectories for construction of the reduced basis. 
As initial basis we set $N_0=1$ and $ V_{N_0}:= P_0$ using the parameter independent initial condition.

The POD-Greedy algorithm produces a basis of 33 vectors and the 
overall computation of the reduced basis takes 7.9 hours, 
the dominating computation time being spent in the error evaluations 
and POD computations. Some of the resulting orthonormal basis vectors 
are illustrated in Figure \ref{fig:basis-vectors}.
The error decay curve and the selected parameters in the parameter domain are 
illustrated in Figure 
\ref{fig:POD-Greedy-details}. We nicely observe an exponential error decay, which indicates 
a parametric smoothness of the solution manifold, cf.\ the convergence 
rate statement given before for the POD-Greedy algorithm.
The selected parameters seem to be located at the boundary of the parameter domain, indicating that the model behaviour in between can well be interpolated from the model behaviours along the boundary of the parameter domain.
%So a refined procedure could make use of this by an adopted training set.

The final reduced model of dimension 33 can then be simulated in 
0.135 seconds, corresponding to a computational speedup factor of more 
than 700.

\begin{figure}
  \centering
  \includegraphics{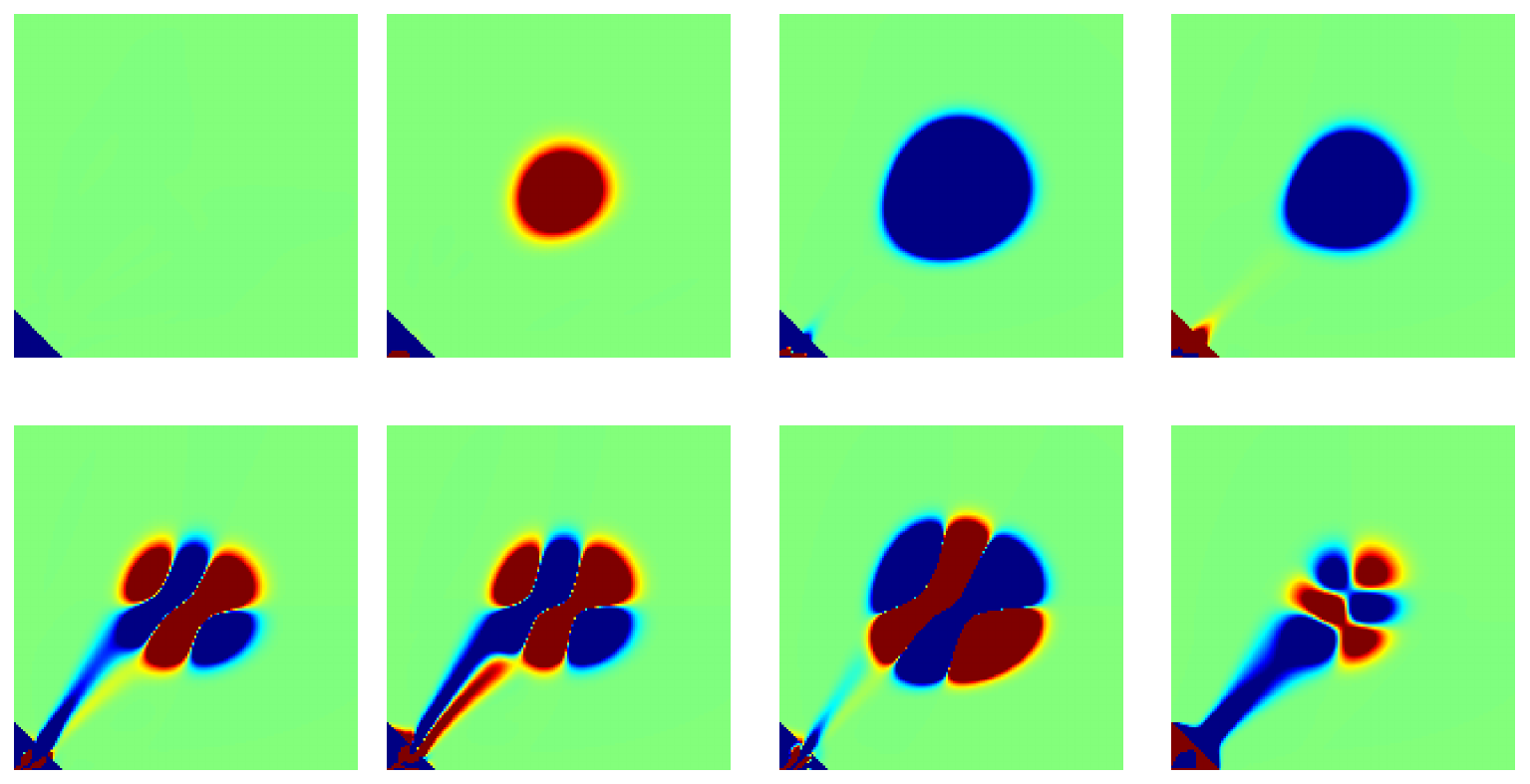}
  \caption{Illustration of the first eight basis vectors for the switch model generated by the 
POD-Greedy algorithm.}
\label{fig:basis-vectors}
\end{figure}

\begin{figure}
  \centering
  \includegraphics{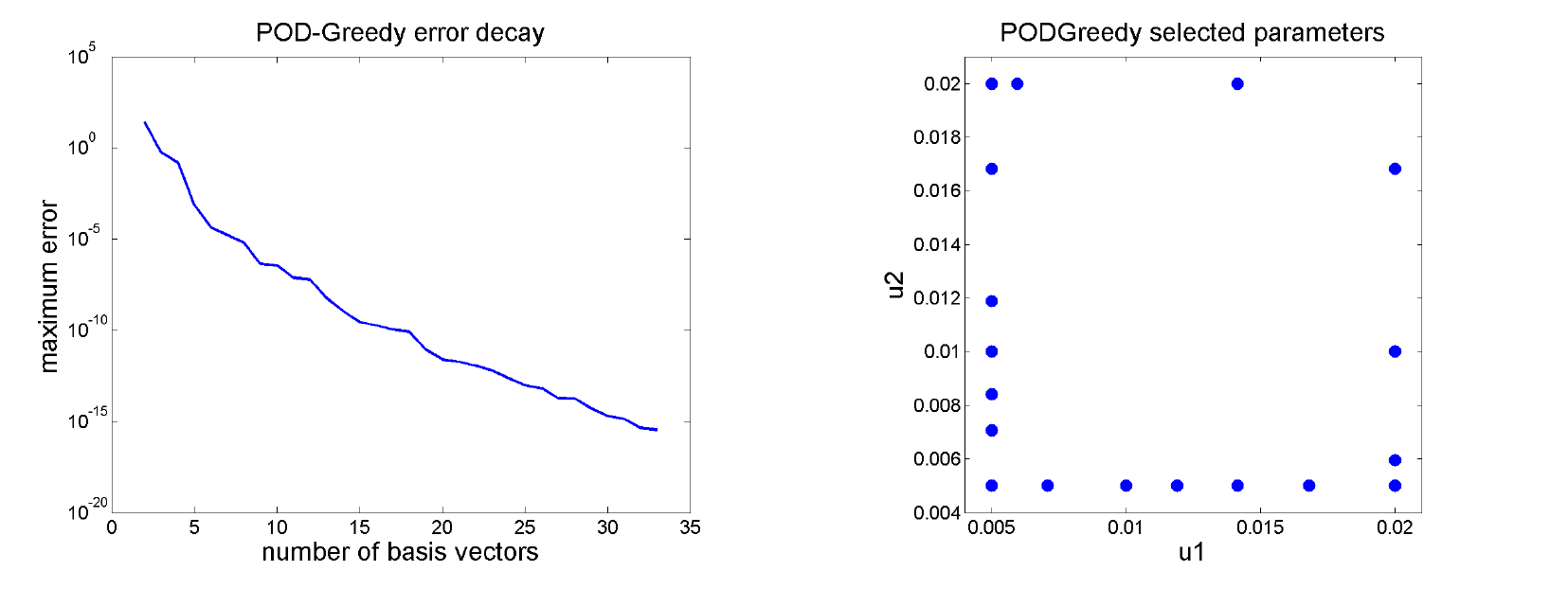}
  \caption{Illustration of the error decay during the POD-Greedy 
algorithm (right) applied to the switch model and the selected parameters (left) being a 
small subset of the 81 training parameter points.}
\label{fig:POD-Greedy-details}
\end{figure}

\subsubsection*{Parameter estimation}

We exemplify a possible application of the reduced order model in parameter estimation,
where we assume that a distorted output $y(t)$ as the expected 
values $E[x_1]$ is available from
population-averaged measurements.
The task is to estimate the parameter values $u_1$ and $u_2$ from such a 
noisy measurement.

The reference parameter is $\param_{ref} = (u_1,u_2) = (0.01,0.01)^T$, and, for the
purpose of this example, the measured output is produced by simulating the original
model with the reference parameter values
and adding 5\% relative random white noise $n(t)$ sampled from a standard 
normal distribution, $y_{meas}(t):=y(t,\param_{ref})(1 + 0.05 n(t))$.
An illustration of the reference output $y(t,\param_{ref})$ and the noisy signal $y_{meas}(t)$ is given in the left of Figure~\ref{fig:optimization-surface}.
 
\begin{figure}
  \centering
  \includegraphics{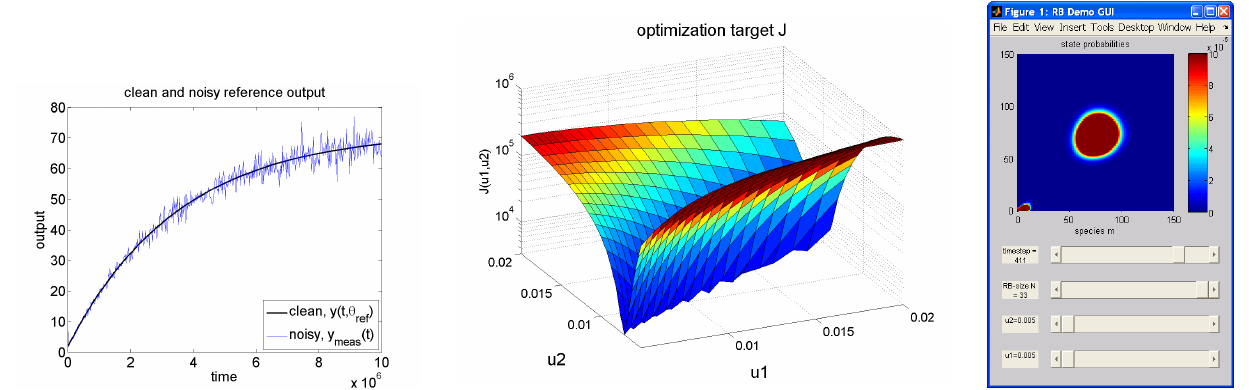}
  \caption{Parametric analysis results for the reduced switch model:
Application of parametric reduced models for parametric analysis:
Illustration of the clean and noisy signals 
$y(t,\param_{ref})$ and $y_{meas}(t)$, respectively (left), the optimization 
target $J(\param)$ over the parameter domain (middle),
interactive parameter exploration by
a graphical user interface (right).}
\label{fig:optimization-surface}
\end{figure}

We want to recover the values of the parameters $u_1$ and $u_2$ based 
on fitting the reduced parametric model's output $\hat y(t,\param)$ 
to the measured output $y_{meas}(t)$.
As is commonly done in parameter estimation, we formulate a least squares cost function as
\begin{equation} \label{eqn:optim-target}
\begin{aligned}
J(\param) = \int_0^T (y_{meas}(t) - \hat y(t,\param))^2 dt,
\end{aligned}
\end{equation}
and estimate the parameters by
\begin{equation}
  \label{eq:optim-problem}
  \param_{est} = \arg \min_{\param \in \cP} J(\param).
\end{equation}

In such an optimization problem, typically many forward simulations are 
required for adjusting $\hat y$ to the measurement. This is a particular 
beneficial scenario for reduced order models, as these simulations can be 
computed rapidly. 

In order to gain a deeper insight into the optimization problem~\eqref{eq:optim-problem},
we plot the values of the error functional $J(\param)$ over the parameter 
domain (middle of Figure~\ref{fig:optimization-surface}).
Using the reduced model, the computation of the required $21^2=441$ trajectories is realized in 
less than one minute. 
This would be a significant computational effort when using a 
non-reduced model.

From the cost function plot, we observe a narrow area of parameters which seem to produce a similar output 
as the reference parameter $\param_{ref}$.
This shows that the two model parameters are not simultaneously identifiable from the considered output,
and indicates that there may exist a functional dependence between the 
parameters $u_1$ and $u_2$ such that the model yields similar outputs $y(t)$.

Assuming a functional dependence of $u_1$ and $u_2$ we now consider the 
1-dimensional optimization problem along the line $u_2 = u_{2,ref} = 0.01$.
We would like to recover $u_1$ from the optimization problem. 
The corresponding value of the cost function
is $J(\param_{ref})= 3330.68$, indicating a significant contribution of the noise.
This restricted optimization problem is well conditioned and
the optimization with a standard active set algorithm by MATLAB's 
command {\tt fmincon} yields the estimated parameter 
$\param_{est}:=(u_{1,est},0.01)$ with
$u_{1,est} = 0.0100204$,
using 27 evaluations of the cost function.
This accounts to a relative error in 
the $u_1$ value of  0.204\%, hence 
excellent recovery.
We refrain from plotting the recovered output $\hat y(t,\param_{est})$ as it 
is visually indiscriminable from the output in the left of Fig.\ 
\ref{fig:optimization-surface}).
Interestingly, the optimization target value  
$J(\param_{est})=3329.56$ implies $J(\param_{est}) < J(\param_{ref})$,
which may stem from a slight approximation error in the reduced model or from the effects of the measurement noise.
% , hence, 
% the reduced model with the estimated variable indeed gives 
% a better fit to the full model's noisy signal than using the 
% reference value in the reduced model.
%, which 
%would be a significant computational effort when using a non-reduced model.

The right plot in Fig. \ref{fig:optimization-surface} illustrates another 
application of reduced parametric models: We incorporated the model in 
an interactive graphical user interface in
{\em RBmatlab}, a matlab package for model order reduction,  
available for download at {\tt www.morepas.org}.
This allows interactive parameter variations and instantaneous 
simulation response.

\subsection*{Sensitivity analysis in a stochastic oscillator}
\label{sec:oscill-model}

\subsubsection*{Network description}
\label{sec:oscill-network-description}

The second case study is built on a genetic oscillator model showing stochastic resonance, which was presented in \cite{El-SamadKha2006a}.
The oscillator is based on a negative feedback loop between two genes with one gene having positive autoregulation.
The oscillator is modelled as a reaction network with two species $X_1$, $X_2$, representing the gene products.
The network reactions are specified in Table~\ref{tab:oscillator-model}.
In the original model in \cite{El-SamadKha2006a}, the dynamics were described as stochastic differential equation for the amounts of $X_1$ and $X_2$, coming from a Langevin approximation to the stochastic dynamics \cite{Higham2008}.
For the framework used in this paper, the dynamics have to be described directly by the underlying CME.
To achieve this, we introduce the parameter $s$ which maps the dimensionless state variables from \cite{El-SamadKha2006a} to actual molecule numbers as required for the CME.
Thus, $s$ is also a measure for the network's noise level: the higher $s$, the larger the molecule number that is considered, and the smaller the noise level will be.

\begin{table}
  \centering
  \caption{List of reactions and reaction propensity functions for the oscillator model adopted from \cite{El-SamadKha2006a}.
Nominal parameter values are
    $k_1 = 15 \frac{1}{\mathrm{s}}$, 
    $k_2 = 0.2$,
    $k_3 = 1 \frac{1}{\mathrm{s}}$, 
    $k_4 = 10 \frac{1}{\mathrm{s}}$, 
    $k_5 = 100 \frac{1}{\mathrm{s}}$, 
    $k_6 = 6.5$,
    $k_7 = 100 \frac{1}{\mathrm{s}}$, 
    $s = 10$.}
  \label{tab:oscillator-model}
  \begin{tabular}{ccc}
    Reaction & Stoichiometry $v_j$ & Propensity $\propensity_j$ \\
    Production of $X_1$ & $(1, 0)\T$ & $\frac{k_1 s^2}{k_2 s + x_2}$ \\
    Degradation of $X_1$ & $(-1, 0)\T$ & $k_3 x_1$ \\
    Production of $X_2$ & $(0, 1)\T$ & $k_4 s + \frac{k_5 x_2^2 x_1}{k_6 s^2 + x_2^2}$ \\
    Degradation of $X_2$ & $(0, -1)\T$ & $k_7 x_2$ \\
  \end{tabular}
\end{table}

The network model in Table~\ref{tab:oscillator-model} shows oscillations only in a stochastic description.
The deterministic model has a unique asymptotically stable equilibrium point, but in a stochastic model, fluctuations may push the molecular numbers beyond a certain threshold, inducing a dynamical response along a slow mani\-fold, which corresponds to one oscillatory period \cite{El-SamadKha2006a}.
Depending on the noise level, such responses will be initiated more or less often, corresponding to a more or less regular oscillatory pattern.

The system is truncated to the rectangle
$\bar \Omega:= \{0,\ldots,300\}\times\{0,\ldots,300\}$, which contains the relevant system states for the parameter ranges of interest.

Similarly as in the switch example, the reaction propensity expressions contain rational terms in the parameters $s$, $k_2$, and $k_6$.
These three cannot be decomposed directly, so we do the decomposition described in the methods section for the other five parameters only.
With this decomposition, the truncated CME for the genetic oscillator can be written as
\begin{equation}
  \label{eq:oscill-detailed-ode}
  \dot P(t) = \bigl(k_1 A^{[1]} + k_3 A^{[2]} + k_4 A^{[3]} + k_5 A^{[4]} + k_7 A^{[5]}\bigr) P(t),
\end{equation}
where $A^{[i]}$, $i=1,\dotsc,5$ are of dimension $301^2 \times 301^2 = 90601 \times 90601$.
The initial condition for \eqref{eq:oscill-detailed-ode} is chosen as a uniform distribution over the rectangle $\{0, \ldots, 50\} \times \{0, \ldots, 50\}$:
\begin{equation}
  \label{eq:oscill-ic}
  p(0,x) = \left\lbrace
    \begin{aligned}
      \frac{1}{51^2} 
    &\quad \textnormal{for } x_1 \leq 50,\ x_2 \leq 50 \\
      0 &\quad \textnormal{otherwise.}
    \end{aligned}\right.
  \end{equation}
The time scale of interest for the model in \eqref{eq:oscill-detailed-ode} is for $0 \leq t \leq T = 6$.
At the end of the interval, the probability distribution seems to approach a steady state.

Some state plots are given in Figure \ref{fig:oscill-state-plots}. 
One observes a significant effect of the parameter $k_4$ on the amplitude of the oscillations.
The simulation time for the detailed model was in average 7.3 minutes on a 
Dell desktop computer with 3.2 GHz dual-core Intel 4 processor and 1 GB RAM, 
without including the computation time for the construction of the state 
transition matrix $A(\param)$.

\begin{figure}
  \centering
  \includegraphics{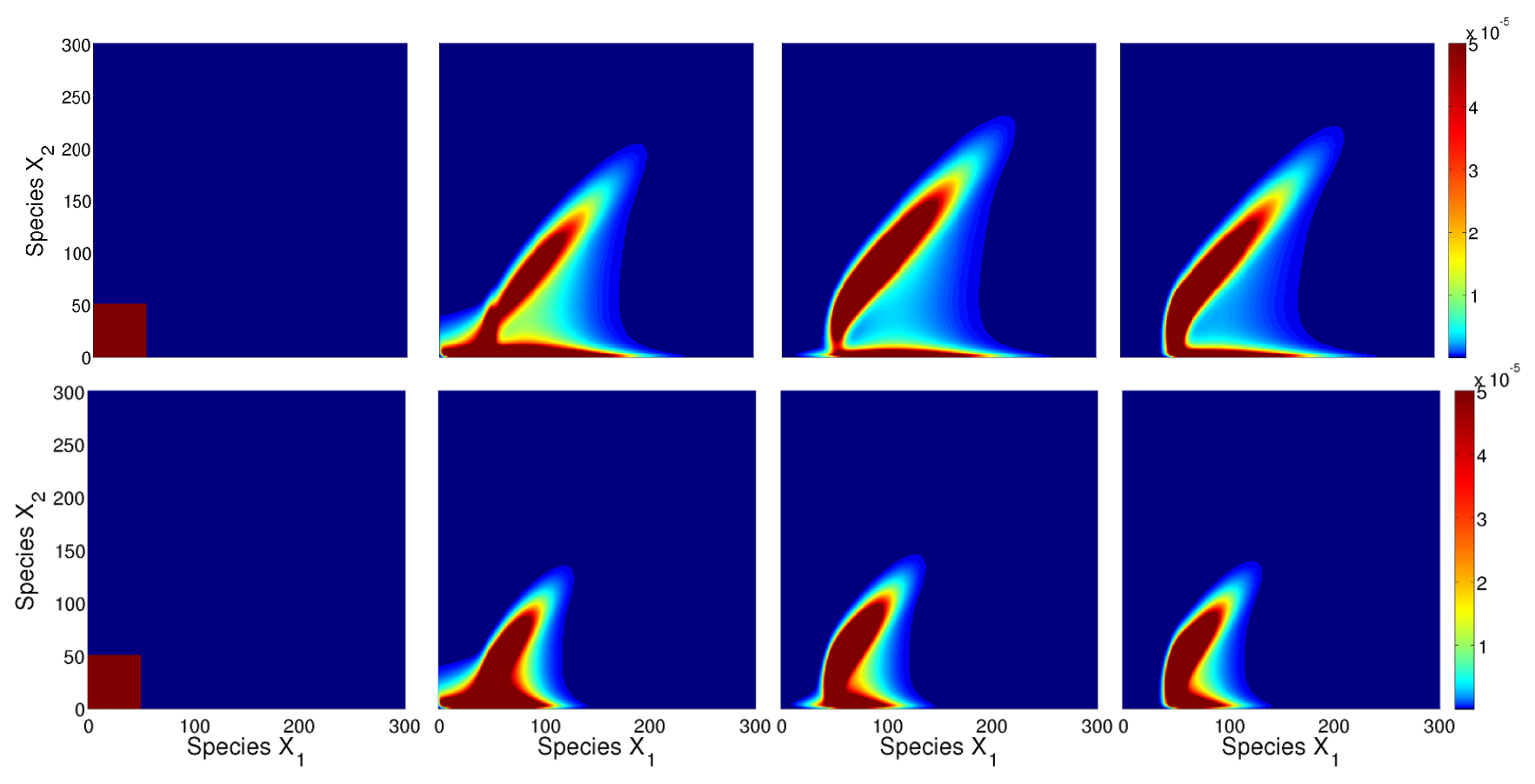}
  \caption{Illustration of some solution snapshots $P(t)$ of the oscillator CME model for parameter values
$k_4 = 15$ (upper row) and
$k_4 = 30$ (lower row) at times
$t = 0,\ 0.2,\ 0.6,\ 6.0$ from left to right.}
\label{fig:oscill-state-plots}
\end{figure}

\subsubsection*{Basis generation}
\label{sec:oscill-basis-cme}

% Generated RB basis on fixed M_train with 109 basis vectors.
% Runtime = 59242.492
% computation time for POD Greedy:

% pod_time =

%    5.9244e+04
For the basis generation, the parameter $k_4$ was assumed to vary within the interval $[10,\ 100]$.
A reduced basis with the POD-Greedy algorithm was computed from a training set of 30 logarithmically
equidistant parameters over the parameter domain (Figure~\ref{fig:oscillator_sensitivity}).
As in the switch example,
the target accuracy was chosen as $\varepsilon_{tol}= 10^{-12}$,
and the initial basis was chosen from the initial condition $V_{1} := P_0$.

The POD-Greedy algorithm produces a basis of 109 vectors, with an 
overall computation time of 16.5 hours on the hardware as in the 
previous subsection.
%a Dell desktop computer with 3.2 GHz dual-core 
% Intel 4 processor and 1 GB RAM.
The first 20 basis vectors are shown in Figure~\ref{fig:oscillator-basis}.
It is apparent that several of the basis vectors are directly included in order to reproduce the different amplitudes of oscillations that will occur under variations of the parameter $k_4$.
The error decay curve is shown in Figure~\ref{fig:oscillator-error-decay}, displaying an exponential error decay as also observed for the switch example.

\begin{figure}
  \centering
  \includegraphics{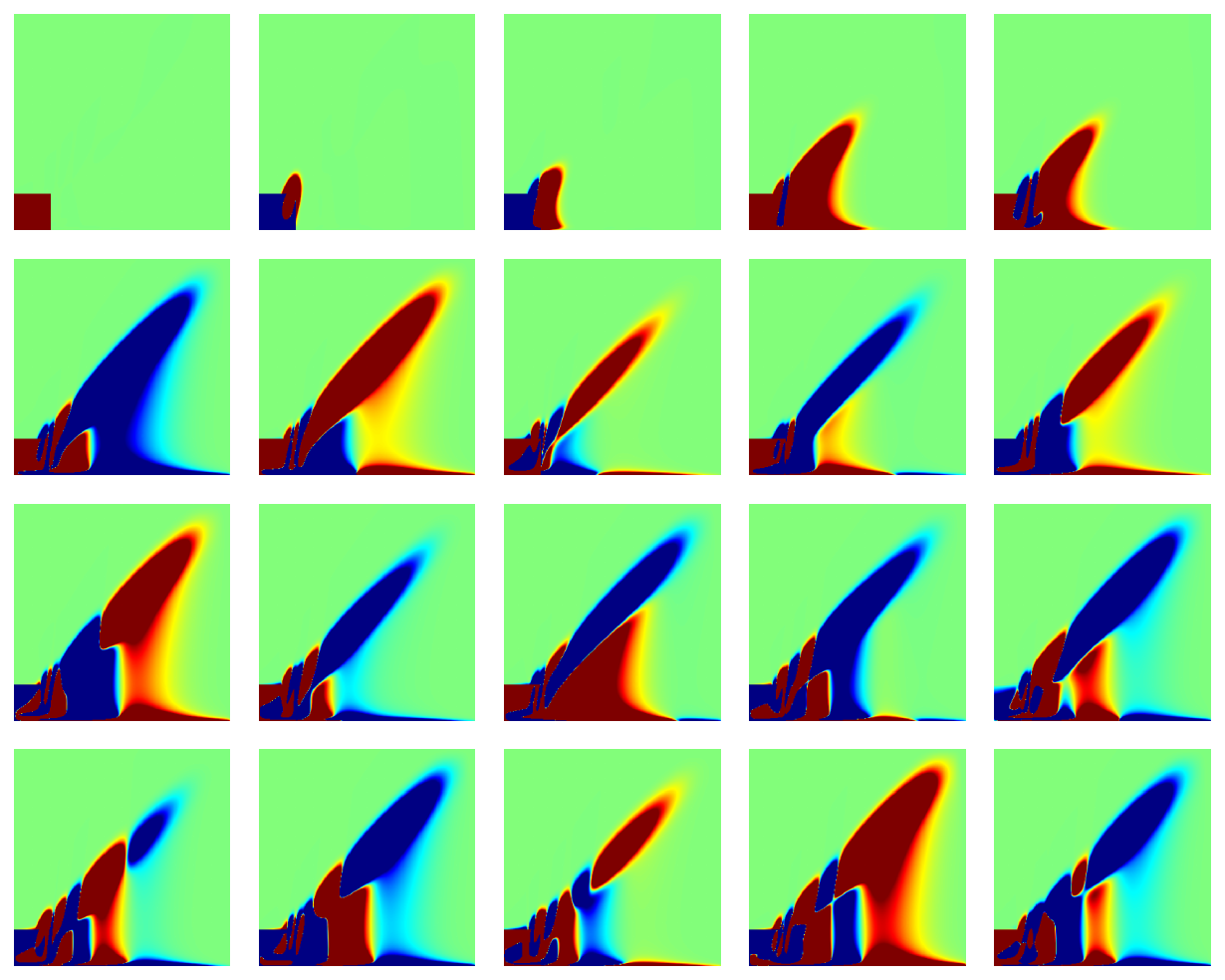}
  \caption{First 20 basis vectors for the oscillator model}
\label{fig:oscillator-basis}
\end{figure}

\begin{figure}
  \centering
  \includegraphics[width=7cm]{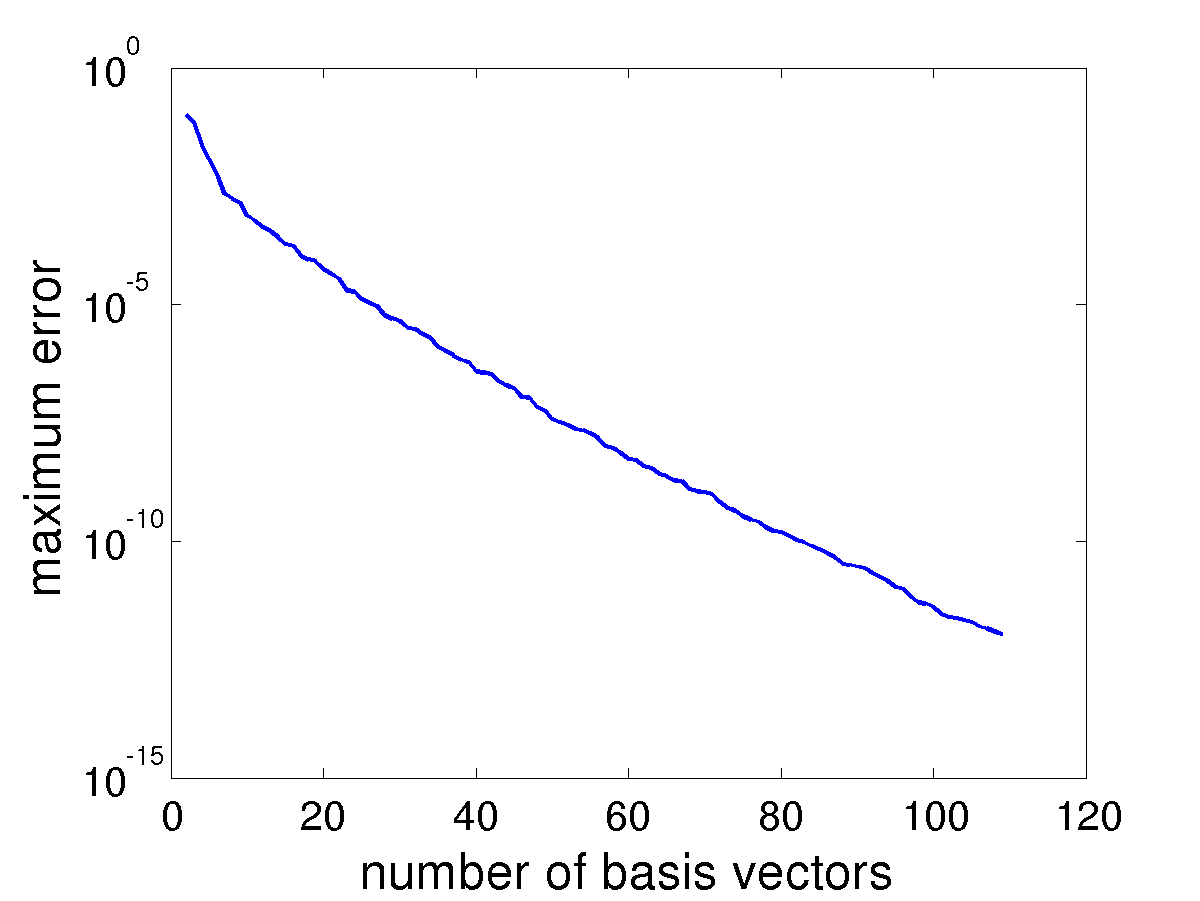}
  \caption{Results of the POD-Greedy algorithm for the oscillator model:
Error decay curve for the oscillator model}
\label{fig:oscillator-error-decay}
\end{figure}

With the reduced basis $V \in \Real^{90601 \times 109}$, we can construct a reduced parametric model for the CME of the oscillator as
\begin{equation}
  \label{eq:reduced-oscillator-model}
  \begin{aligned}
    \dot P_r(t) &= (k_4 A^{[3]}_r + A^{[o]}_r ) P_r(t) \\
    P_{r}(0) &= V\T P(0),
  \end{aligned}
\end{equation}
with $A^{[3]}_r = V\T A^{[3]} V \in \Real^{109\times 109}$ and $A^{[o]}_r = V\T \bigl(k_1 A^{[1]} + k_3 A^{[2]} + k_5 A^{[4]} + k_7 A^{[5]}\bigr) V \in \Real^{109 \times 109}$.
Note that since only $k_4$ has been varied in the reduction process, the other parameters are no longer present as parameters in the reduced model, but just take their nominal values.
While the same basis $V$ could be used to construct another reduced model where all parameters are retained, it is unlikely that this other model will be a good approximation of the original one for varying values of the other parameters.

\subsubsection*{Sensitivity analysis of the oscillation amplitude}

As an application of the reduced order parametric model obtained in the previous section, we study the variations of oscillatory amplitude over a parameter range.
Specifically, we consider 200 equally spaced values for the parameter $k_4$ in the interval $[12,\ 40]$ and compute the probability that the amount of $X_2$ is larger than 100:
\begin{equation}
  \label{eq:oscill-amplitude}
  \mathrm{Prob}(x_2 > 100) = \sum_{x : x_2 > 100} p(T, x),
\end{equation}
with $T = 6$ the final time of the simulation.
The results are shown in Figure~\ref{fig:oscillator_sensitivity} and show a clear decay of oscillatory amplitude for increasing values of $k_4$.
% TODO: give computation time with the reduced model.
Due to the significant time savings from the reduced model, this sensitivity curve can be computed with a high resolution.

\begin{figure}
  \centering
  \includegraphics{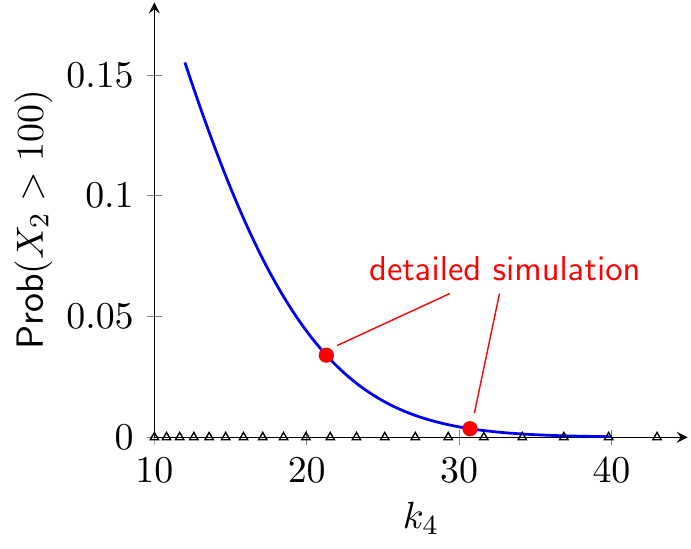}
  \caption{Sensitivity analysis of oscillation amplitude over a parameter interval.
Blue line shows oscillatory amplitude over the parameter $k_4$ predicted from the reduced model.
Red dots are validation results from a simulation of the original model.
Triangles on the parameter axis indicate parameter values which were used in the construction of the reduced basis.}
\label{fig:oscillator_sensitivity}
\end{figure}

To evaluate the quality of the reduced model, we also computed the probability~\eqref{eq:oscill-amplitude} using the original model \eqref{eq:oscill-detailed-ode} at two points within the considered interval for the parameter $k_4$.
As shown in Figure~\ref{fig:oscillator_sensitivity}, the results from the original model are in perfect agreement with the predictions from the reduced model at these points.
Since the points at which the original model was evaluated in this experiment were not part of the training set (shown as triangles on the parameter axis in Figure~\ref{fig:oscillator_sensitivity}), this shows that it is in fact possible to extrapolate the reduced model to parameter values that were not used to construct the basis.

\section*{Conclusions}
\label{sec:conclusion}

In this paper, we have introduced the application of parametric model reduction methods to finite-state approximations of the chemical master equation.
We have also presented two case studies where these methods are applied to CME models of different networks in order to make parametric analysis tasks computationally efficient.
By this, it has become clear that parametric model reduction methods are a very useful tool for the analysis of stochastic biochemical reaction network described by the CME.

Especially analysis tasks where many repeated simulations of a network with different parameter values are required can profit significantly from parametric model reduction.
This includes for example sensitivity analysis or parameter optimization tasks such as identifiability analysis or estimation.
Moreover, the significant speedup of the simulation for the reduced model allows an interactive exploration of the network's dynamics within the parameter space within a suitable graphical user interface.

This contribution is just a first step in the application of parametric model reduction methods to the CME.
One particularly important aspect that we have not discussed here is the computation of error estimates for certifying that the simulation output of the reduced model is within some tolerance of the corresponding simulation output of the original model.
To maintain computational efficiency, the error estimation should be done without actually simulating the original model.
Error estimation methods have been developed for parametric model reduction of generic models \cite{HaasdonkOhlberger2010}, but tighter estimates could likely be obtained by taking into account the special structure of the CME models.
Recent work for example refined the previous generic error bounds for stable models \cite{HasenauerLoh2012}.

% --------------------------------------------------------------
% -------------- Acknowledgements -------------------------------
% --------------------------------------------------------------

\section*{Authors contributions}
SW and BH conceived of the study, performed the study, and wrote the manuscript.
Both authors read and approved the final manuscript.

\section*{Acknowledgements}
  \ifthenelse{\boolean{publ}}{\small}{}
We thank Wolfgang Halter for programming support in the oscillator case study.
The authors would like to thank the German Research Foundation (DFG) for financial support of the project within the Cluster of Excellence in Simulation 
Technology at the University of Stuttgart.
BH also acknowledges the Baden-W\"urttemberg Stiftung gGmbH for funding.
This work was also supported by the German Research Foundation (DFG) within the funding programme Open Access Publishing.

{\ifthenelse{\boolean{publ}}{\footnotesize}{\small}
 % \bibliographystyle{bmc_article}  % Style BST file
 % \bibliography{refs-cmemodred}    % Bibliography file (usually '*.bib' )
 %% BioMed_Central_Bib_Style_v1.01

\newcommand{\BMCxmlcomment}[1]{}

\BMCxmlcomment{

<refgrp>

<bibl id="B1">
  <title><p>A rigorous derivation of the chemical master equation</p></title>
  <aug>
    <au><snm>Gillespie</snm><fnm>D. T.</fnm></au>
  </aug>
  <source>Physica A: Statist. Theor. Phys.</source>
  <pubdate>1992</pubdate>
  <volume>188</volume>
  <issue>1-3</issue>
  <fpage>404</fpage>
  <lpage>-425</lpage>
  <url>http://www.sciencedirect.com/science/article/B6TVG-46FX396-7N/2/a0537c1efc0f5c330fa05b5e4ae61b98</url>
</bibl>

<bibl id="B2">
  <title><p>Stochastic processes in physics and chemistry</p></title>
  <aug>
    <au><snm>Kampen</snm><fnm>N. G.</fnm></au>
  </aug>
  <publisher>North-Holland Amsterdam</publisher>
  <pubdate>1981</pubdate>
</bibl>

<bibl id="B3">
  <title><p>The finite state projection algorithm for the solution of the
  chemical master equation</p></title>
  <aug>
    <au><snm>Munsky</snm><fnm>B.</fnm></au>
    <au><snm>Khammash</snm><fnm>M.</fnm></au>
  </aug>
  <source>J. Chem. Phys.</source>
  <pubdate>2006</pubdate>
  <volume>124</volume>
  <issue>4</issue>
  <fpage>044104</fpage>
  <url>http://dx.doi.org/10.1063/1.2145882</url>
</bibl>

<bibl id="B4">
  <title><p>A Dynamical Low-Rank Approach to the Chemical Master
  Equation</p></title>
  <aug>
    <au><snm>Jahnke</snm><fnm>T</fnm></au>
    <au><snm>Huisinga</snm><fnm>W</fnm></au>
  </aug>
  <source>Bull.\ Math.\ Biol.</source>
  <pubdate>2008</pubdate>
  <volume>70</volume>
  <fpage>2283</fpage>
  <lpage>2302</lpage>
</bibl>

<bibl id="B5">
  <title><p>Sparse grids and hybrid methods for the chemical master
  equation</p></title>
  <aug>
    <au><snm>Hegland</snm><fnm>M.</fnm></au>
    <au><snm>Hellander</snm><fnm>A.</fnm></au>
    <au><snm>L{\"o}tstedt</snm><fnm>P.</fnm></au>
  </aug>
  <source>BIT Numerical Mathematics</source>
  <pubdate>2008</pubdate>
  <volume>48</volume>
  <fpage>265</fpage>
  <lpage>-283</lpage>
</bibl>

<bibl id="B6">
  <title><p>Approximation of Large-Scale Dynamical Systems</p></title>
  <aug>
    <au><snm>Antoulas</snm><fnm>A. C.</fnm></au>
  </aug>
  <publisher>Philadelphia, {USA}: SIAM</publisher>
  <pubdate>2005</pubdate>
</bibl>

<bibl id="B7">
  <title><p>The Finite State Projection Approach for the Analysis of Stochastic
  Noise in Gene Networks</p></title>
  <aug>
    <au><snm>Munsky</snm><fnm>B.</fnm></au>
    <au><snm>Khammash</snm><fnm>M.</fnm></au>
  </aug>
  <source>Automatic Control, IEEE Transactions on</source>
  <pubdate>2008</pubdate>
  <volume>53</volume>
  <issue>Special Issue</issue>
  <fpage>201</fpage>
  <lpage>-214</lpage>
</bibl>

<bibl id="B8">
  <title><p>Parametrische {M}odellreduktion mit d{\"u}nnen
  {G}ittern</p></title>
  <aug>
    <au><snm>Baur</snm><fnm>U.</fnm></au>
    <au><snm>Benner</snm><fnm>P.</fnm></au>
  </aug>
  <source>GMA-Fachausschuss 1.30, Modellbildung, Identifizierung und Simulation
  in der Automatisierungstechnik, Salzburg ISBN 978-3-9502451-3-4</source>
  <pubdate>2008</pubdate>
  <fpage>262</fpage>
  <lpage>-271</lpage>
</bibl>

<bibl id="B9">
  <title><p>Efficient Reduced Models and A-Posteriori Error Estimation for
  Parametrized Dynamical Systems by Offline/Online Decomposition</p></title>
  <aug>
    <au><snm>Haasdonk</snm><fnm>B.</fnm></au>
    <au><snm>Ohlberger</snm><fnm>M.</fnm></au>
  </aug>
  <source>MCMDS, Mathematical and Computer Modelling of Dynamical
  Systems</source>
  <pubdate>2011</pubdate>
  <volume>17</volume>
  <issue>2</issue>
  <fpage>145</fpage>
  <lpage>-161</lpage>
</bibl>

<bibl id="B10">
  <title><p>Multi-parameter moment-matching model-reduction approach for
  generating geometrically parameterized interconnect performance
  models</p></title>
  <aug>
    <au><snm>Daniel</snm><fnm>L.</fnm></au>
    <au><snm>Siong</snm><fnm>O.</fnm></au>
    <au><snm>Chay</snm><fnm>L.</fnm></au>
    <au><snm>Lee</snm><fnm>K.</fnm></au>
    <au><snm>White</snm><fnm>J.</fnm></au>
  </aug>
  <source>IEEE Transactions on Computer-Aided Design of Integrated Circuits and
  Systems</source>
  <pubdate>2004</pubdate>
  <volume>23</volume>
  <issue>5</issue>
  <fpage>678</fpage>
  <lpage>-693</lpage>
</bibl>

<bibl id="B11">
  <title><p>Parameter Preserving Model Order Reduction of a Flow
  Meter</p></title>
  <aug>
    <au><snm>Moosmann</snm><fnm>C.</fnm></au>
    <au><snm>Rudnyi</snm><fnm>E.B.</fnm></au>
    <au><snm>Greiner</snm><fnm>A.</fnm></au>
    <au><snm>Korvink</snm><fnm>J.G.</fnm></au>
    <au><snm>Hornung</snm><fnm>M.</fnm></au>
  </aug>
  <source>Technical Proceedings of Nanotech 2005</source>
  <pubdate>2005</pubdate>
</bibl>

<bibl id="B12">
  <title><p>Modeling and Simulating Chemical Reactions</p></title>
  <aug>
    <au><snm>Higham</snm><fnm>D. J.</fnm></au>
  </aug>
  <source>SIAM Rev.</source>
  <publisher>SIAM</publisher>
  <pubdate>2008</pubdate>
  <volume>50</volume>
  <issue>2</issue>
  <fpage>347</fpage>
  <lpage>-368</lpage>
  <url>http://link.aip.org/link/?SIR/50/347/1</url>
</bibl>

<bibl id="B13">
  <title><p>Exact stochastic simulation of coupled chemical
  reactions</p></title>
  <aug>
    <au><snm>Gillespie</snm><fnm>D. T.</fnm></au>
  </aug>
  <source>J.\ Phys.\ Chem.</source>
  <pubdate>1977</pubdate>
  <volume>81</volume>
  <issue>25</issue>
  <fpage>2340</fpage>
  <lpage>-2361</lpage>
  <url>http://pubs.acs.org/cgi-bin/abstract.cgi/jpchax/1977/81/i25/f-pdf/f_j100540a008.pdf</url>
</bibl>

<bibl id="B14">
  <title><p>Reduced Basis Method for Finite Volume Approximations of
  Parametrized Linear Evolution Equations</p></title>
  <aug>
    <au><snm>Haasdonk</snm><fnm>B.</fnm></au>
    <au><snm>Ohlberger</snm><fnm>M.</fnm></au>
  </aug>
  <source>M2AN, Math. Model. Numer. Anal.</source>
  <pubdate>2008</pubdate>
  <volume>42</volume>
  <issue>2</issue>
  <fpage>277</fpage>
  <lpage>-302</lpage>
</bibl>

<bibl id="B15">
  <title><p>An hp Certified Reduced Basis Method for Parametrized Parabolic
  Partial Differential Equations</p></title>
  <aug>
    <au><snm>Eftang</snm><fnm>J. L.</fnm></au>
    <au><snm>Knezevic</snm><fnm>D. J.</fnm></au>
    <au><snm>Patera</snm><fnm>A. T.</fnm></au>
  </aug>
  <source>MCMDS, Mathematical and Computer Modelling of Dynamical
  Systems</source>
  <pubdate>2011</pubdate>
  <volume>17</volume>
  <issue>4</issue>
  <fpage>395</fpage>
  <lpage>-422</lpage>
</bibl>

<bibl id="B16">
  <title><p>A Certified Reduced Basis Method for the {F}okker-{P}lanck Equation
  of Dilute Polymeric Fluids: {FENE} Dumbbells in Extensional Flow.</p></title>
  <aug>
    <au><snm>Knezevic</snm><fnm>D.J.</fnm></au>
    <au><snm>Patera</snm><fnm>A.T.</fnm></au>
  </aug>
  <source>SIAM Journal of Scientific Computing</source>
  <pubdate>2010</pubdate>
  <volume>32</volume>
  <issue>2</issue>
  <fpage>793</fpage>
  <lpage>-817</lpage>
</bibl>

<bibl id="B17">
  <title><p>Model {R}eduction using {P}roper {O}rthogonal
  {D}ecomposition</p></title>
  <aug>
    <au><snm>Volkwein</snm><fnm>S.</fnm></au>
  </aug>
  <pubdate>2011</pubdate>
  <url>http://www.uni-graz.at/imawww/volkwein/publist.html</url>
  <note>Lecture Notes, University of Constance</note>
</bibl>

<bibl id="B18">
  <title><p>Principal Component Analysis</p></title>
  <aug>
    <au><snm>Joliffe</snm><fnm>I.T.</fnm></au>
  </aug>
  <publisher>John Wiley \& Sons</publisher>
  <pubdate>2002</pubdate>
</bibl>

<bibl id="B19">
  <title><p>Convergence Rates of the {POD}-{G}reedy Method</p></title>
  <aug>
    <au><snm>Haasdonk</snm><fnm>B.</fnm></au>
  </aug>
  <source>SimTech Preprint 2011-23</source>
  <pubdate>2011</pubdate>
</bibl>

<bibl id="B20">
  <title><p>A Training Set and Multiple Bases Generation Approach for
  Parametrized Model Reduction Based on Adaptive Grids in Parameter
  Space.</p></title>
  <aug>
    <au><snm>Haasdonk</snm><fnm>B.</fnm></au>
    <au><snm>Dihlmann</snm><fnm>M.</fnm></au>
    <au><snm>Ohlberger</snm><fnm>M.</fnm></au>
  </aug>
  <source>MCMDS, Mathematical and Computer Modelling of Dynamical
  Systems</source>
  <pubdate>2011</pubdate>
  <volume>17</volume>
  <issue>4</issue>
  <fpage>423</fpage>
  <lpage>442</lpage>
</bibl>

<bibl id="B21">
  <title><p>Constructing stochastic models from deterministic process equations
  by propensity adjustment.</p></title>
  <aug>
    <au><snm>Wu</snm><fnm>J.</fnm></au>
    <au><snm>Vidakovic</snm><fnm>B.</fnm></au>
    <au><snm>Voit</snm><fnm>E. O</fnm></au>
  </aug>
  <source>BMC Syst Biol</source>
  <pubdate>2011</pubdate>
  <volume>5</volume>
  <issue>1</issue>
  <fpage>187</fpage>
  <url>http://dx.doi.org/10.1186/1752-0509-5-187</url>
</bibl>

<bibl id="B22">
  <title><p>An 'empirical interpolation' method: application to efficient
  reduced-basis discretization of partial differential equations</p></title>
  <aug>
    <au><snm>Barrault</snm><fnm>M.</fnm></au>
    <au><snm>Maday</snm><fnm>Y.</fnm></au>
    <au><snm>Nguyen</snm><fnm>N.C.</fnm></au>
    <au><snm>Patera</snm><fnm>A.T.</fnm></au>
  </aug>
  <source>C. R. Math. Acad. Sci. {P}aris {S}eries {I}</source>
  <pubdate>2004</pubdate>
  <volume>339</volume>
  <fpage>667</fpage>
  <lpage>-672</lpage>
</bibl>

<bibl id="B23">
  <title><p>Bridging time scales in cellular decision making with a stochastic
  bistable switch</p></title>
  <aug>
    <au><snm>Waldherr</snm><fnm>S.</fnm></au>
    <au><snm>Wu</snm><fnm>J.</fnm></au>
    <au><snm>Allg{\"o}wer</snm><fnm>F.</fnm></au>
  </aug>
  <source>BMC Syst.\ Biol.</source>
  <pubdate>2010</pubdate>
  <volume>4</volume>
  <fpage>108</fpage>
  <url>http://www.biomedcentral.com/1752-0509/4/108</url>
</bibl>

<bibl id="B24">
  <title><p>Coherence resonance: a mechanism for noise induced stable
  oscillations in gene regulatory networks</p></title>
  <aug>
    <au><snm>{El-Samad}</snm><fnm>H.</fnm></au>
    <au><snm>Khammash</snm><fnm>M.</fnm></au>
  </aug>
  <source>Proc.\ of the 45th Conf.\ Dec.\ Contr.\ (CDC), San Diego,
  USA</source>
  <pubdate>2006</pubdate>
  <fpage>2382</fpage>
  <lpage>-2387</lpage>
</bibl>

<bibl id="B25">
  <title><p>Dynamical optimization using reduced order models: A method to
  guarantee performance</p></title>
  <aug>
    <au><snm>Hasenauer</snm><fnm>J</fnm></au>
    <au><snm>L{\"o}hning</snm><fnm>M</fnm></au>
    <au><snm>Khammash</snm><fnm>M</fnm></au>
    <au><snm>Allg{\"o}wer</snm><fnm>F</fnm></au>
  </aug>
  <pubdate>2012</pubdate>
  <note>Journal of Process Control, Online Publication before print</note>
</bibl>

</refgrp>
} % end of \BMCxmlcomment

} 

%%%%%%%%%%%

\ifthenelse{\boolean{publ}}{\end{multicols}}{}

\end{bmcformat}
\end{document}